\documentclass[12pt]{article}
\usepackage{docmute}

\usepackage{cite}
\usepackage{graphicx}
\usepackage{dcolumn}
\usepackage{bm}
\usepackage[hidelinks]{hyperref}

\usepackage{epsf}
\usepackage{amsmath,amssymb}
\usepackage{slashed}
\usepackage{graphicx}
\usepackage{subfig}

\renewcommand{\thefootnote}{\fnsymbol{footnote}}
\def\thefootnote{\fnsymbol{footnote}}
\usepackage{comment}

\usepackage{tikz}
\usetikzlibrary{quantikz2}

\begin{document}

\renewcommand{\thefootnote}{\fnsymbol{footnote}}
\setcounter{footnote}{0}

\begin{titlepage}

\def\thefootnote{\fnsymbol{footnote}}

\begin{center}

\hfill July, 2024\\

\vskip .5in

{\Large \bf
  
  Search for QCD axion dark matter with transmon qubits and quantum circuit

}

\vskip .5in

{\large
  Shion Chen$^{(a)}$,
  Hajime Fukuda$^{(b)}$,
  Toshiaki Inada$^{(c)}$,
  Takeo Moroi$^{(b,d)}$,
  Tatsumi Nitta$^{(c)}$,
  Thanaporn Sichanugrist$^{(b)}$\footnote{Corresponding author}
}

\vskip .5in

$^{(a)}$
{\em Department of Physics, Kyoto University, Kitashirakawa-Oiwakecho, Sakyo-ku, Kyoto 606-8502, Japan}

\vskip 0.1in

$^{(b)}$
{\em Department of Physics, The University of Tokyo, 7-3-1 Hongo, Bunkyo-ku, Tokyo 113-0033, Japan}

\vskip 0.1in

$^{(c)}$
{\em International Center for Elementary Particle Physics (ICEPP), The University of Tokyo, 7-3-1 Hongo, Bunkyo-ku, Tokyo 113-0033, Japan}

\vskip 0.1in

$^{(d)}$
{\em QUP (WPI), KEK, Oho 1-1, Tsukuba, Ibaraki 305-0801, Japan}

\end{center}
\vskip .5in

\begin{abstract}

We propose a direct axion dark matter (DM) search using superconducting transmon qubits as quantum sensors. With an external magnetic field applied, axion DM generates an oscillating electric field which causes the excitation of the qubit; such an excitation can be regarded as a signal of the axion DM. We provide a theoretical consideration of the excitation process of the qubits taking into account the effects of the shielding cavity surrounding the qubits and estimate the signal rate for the axion DM detection. We also discuss the enhancement of the DM signal by using cavity resonance and entangled quantum sensors realized by a quantum circuit. Combining these two effects, we can reach the parameter region suggested by QCD axion models.

\end{abstract}

\end{titlepage}

\renewcommand{\thepage}{\arabic{page}}
\setcounter{page}{1}
\renewcommand{\thefootnote}{\#\arabic{footnote}}
\setcounter{footnote}{0}
\renewcommand{\theequation}{\thesection.\arabic{equation}}

\section{Introduction}
\setcounter{equation}{0}

The use of quantum sensors for direct detection of dark matter (DM) is widely studied \cite{Chou:2023hcc,Kennedy:2020bac,Dixit:2020ymh,Chen:2022quj, Fan:2022uwu,Chigusa:2023hms,Engelhardt:2023qjf,Chen:2023swh,Agrawal:2023umy,Ito:2023zhp,Braggio:2024xed,Das:2022srn,Das:2024jdz,Moretti2024}. The merits of using quantum sensors include not only various interactions that can be expected between quantum sensors and DM but also the evasion of standard quantum limits \cite{Lamoreaux:2013koa}, the realization of high quality factors for the detection, and the enhancement of the DM signal with entangled sensors \cite{doi:10.1126/science.1104149,Chen:2023swh}.
Specifically, a quantum bit (qubit) can be a good DM sensor equipped with its useful properties: controllability of state, precise readout, and tunability of the energy gap. In addition, the qubit technology is growing rapidly with the strong drive from the field of quantum computation~\cite{Kjaergaard2020}.

It has been pointed out in Ref.\ \cite{Chen:2022quj} that transmon qubit~\cite{Koch:2007hay}, a charge-insensitive-type superconducting qubit with good sensitivity to the electromagnetic field, can be a promising quantum sensor to detect wave-like DM \cite{Arias:2012az} which generates effective electric field coupled to electric charges.  Ref.\ \cite{Chen:2022quj} considered the case of the hidden-photon DM and demonstrated that the qubit detector can access the parameter region of the hidden-photon DM which is beyond any current constraints.
This gives a new avenue of high-sensitivity direct detection of DM using a qubit platform to reveal the particle-physics properties of DM, e.g., DM mass and non-gravitational DM interaction with standard-model particles.

The purpose of this paper is to extend the idea and propose the direct detection of the axion DM with transmon qubit utilizing axion-photon interaction as a portal.  We consider the experiment in which a strong static magnetic field is applied around the transmon qubits, with which the axion is converted to the photon; the axion-induced photon may interact with transmon and stimulate the measurable state-transition of transmon. One of the main challenges is the application of a strong magnetic field which is generally disadvantageous for coherence in superconducting devices. However, recent studies~\cite{Krause:2021llk} report that transmon can remain coherent under a static magnetic field as strong as $O(1)$~T when it is perfectly aligned in-plane with respect to the $\sim 30$~nm of thin films of the transmon in contrast to $100-200$~nm of the typical thin film for superconducting qubits.
The ultimate limitation still remains to be explored, but the studies of magnetic resilience of Josephson junction imply that it can go far beyond; for example, a Nb-based ``constriction'' type of Josephson junctions has been reported to survive up to $\sim 5$~T of a magnetic field~\cite{Rokhinson:2012dp}, or Graphene-$\rm{NbSe}_2$ based planar Josephson junctions sustain supercurrent up to 8.5~T in a study of nonreciprocal transport~\cite{Bauriedl2022,Dvir2024}. Transmon operation under strong magnetic fields is also increasingly motivated in the context of studying the Majorana zero modes~\cite{Banerjee2018,_vila_2020}, which may break through the challenge. Note that superconducting qubits working at a high magnetic field have to be well-considered in terms of the critical field of superconductors, the Fraunhofer-type diffraction of Josephson junctions~\cite{Borcsok2019}, and two-level system due to trapped flux~\cite{Graaf2020}. Those potential issues are caused by penetrating flux in the Josephson junctions and superconducting electrodes, hence, a precise alignment system of in-plane magnetic fields by 3-dimensional magnet or piezo actuators and $\sim$ 10~nm thickness of thinner superconducting films has to be developed in a real experiment.

In this paper, we show that, under the $O(1)$ T of a magnetic field applied, a remarkable sensitivity for detecting axion DM is expected in the axion DM search using transmon qubits. Besides, we also discuss in detail the plausible improvement of the DM detection sensitivity with the help of cavity resonance and the quantum entanglement through quantum circuit  \cite{Chen:2023swh}. Combining these two enhancement mechanisms, one can reach the parameter region suggested by QCD axion models such as that of KSVZ~\cite{Kim:1979if, Shifman:1979if} or DFSZ~\cite{Dine:1981rt, Zhitnitsky:1980tq} models.

The organization of the rest of this paper is as follows. In Sec.~\ref{sec:axion&efield}, we elaborate on the nature of axion DM and the electric field converted from axion DM under the presence of an external static magnetic field. In Sec.~\ref{sec:transmonex}, we derive the transmon excitation rate owing to the effect of the axion-induced electric field. We then describe the protocol for using transmon qubit to perform DM direct detection experiments and show the sensitivity we can expect from such a plan in Sec.~\ref{sec:protocol&sen}. We discuss and conclude the result in Sec.~\ref{sec:d&c}.

\section{Axion DM and induced electric field inside cavity} \label{sec:axion&efield}
\setcounter{equation}{0}

We start with overviewing the axion properties. Particularly, we discuss the electric field induced by the axion in the case with an external magnetic field. 

In our discussion, we consider both the QCD axion and axion-like particles as the DM candidates. The QCD axion arises in association with the spontaneous breaking of the Peccei--Quinn symmetry~\cite{Peccei:1977hh,Peccei:1977ur,Weinberg:1977ma,Wilczek:1977pj}, which is introduced to solve the strong CP problem. The QCD axion couples to gauge bosons as well as to fermions in the standard model. The strength of the couplings of the QCD axion as well as its mass are governed by the so-called axion decay constant $f_a$.  On the other hand, the axion-like particles, often abbreviated as ALPs, have more general properties. Particularly, the coupling constants and the mass of the axion-like particles are model-dependent and are often treated as free parameters. In this paper, QCD axions and axion-like particles are collectively called as axions and are denoted as $a$.

In our following discussion, the interaction of the axion with the electromagnetic field plays a crucial role.  The relevant part of the Lagrangian for our discussion is given by
\begin{equation}
    \mathcal{L}= - \frac{1}{4} F_{\mu\nu} F^{\mu\nu} 
    + \frac{1}{2}\partial_\mu a \partial^\mu a - \frac{1}{2} m_a^2 a^2
    - \frac{1}{8} g_{a\gamma \gamma} a \epsilon_{\mu\nu\rho\sigma} F^{\mu\nu} F^{\rho\sigma},
    \label{Lagrangian}
\end{equation}
where $F_{\mu\nu}$ is the field strength of the electromagnetic field and $g_{a\gamma \gamma}$ is a coupling constant.
For the case of the QCD axion, we obtain~\cite{Gorghetto:2018ocs}
\begin{equation}
    m_a\simeq 5.69 \left( \frac{10^{12} \ \mathrm{GeV}}{f_a}\right) \mathrm{\mu eV},
\end{equation}
while the axion-photon coupling can be parameterized as
\begin{equation}
    g_{a\gamma \gamma}= \frac{\alpha}{2\pi} \frac{C_{a\gamma\gamma}}{f_a},
\end{equation}
with $C_{a\gamma\gamma}$ being a model-dependent constant. 
For KSVZ~\cite{Kim:1979if, Shifman:1979if} and DFSZ~\cite{Dine:1981rt, Zhitnitsky:1980tq} models of the QCD axion, the parameter $C_{a\gamma\gamma}$ is given by~\cite{GrillidiCortona:2015jxo}
\begin{equation}
    C^{\rm (KSVZ)}_{a\gamma\gamma} \simeq -1.92,
\end{equation}
and
\begin{equation}
    C^{\rm (DFSZ)}_{a\gamma\gamma} \simeq \frac{8}{3}-1.92,
\end{equation}
respectively.

In our study, we assume that the axion is the dominant component of the DM. We are interested in the axion model in which the axion mass $m_a$ is much smaller than eV. Then, the number of axion particles within the de-Broglie wavelength is so large that the axion can be described as a classical field.  We parameterize the DM axion field as
\begin{equation}
    a(t)=a_0 \cos(m_a t-\alpha),
    \label{a(t)}
\end{equation}
with which the DM energy density is given by
\begin{equation}
  \rho_{\rm DM} = \frac{1}{2} m_a^2 a_0^2.
\end{equation}
Numerically, we adopt the local energy density of DM of
\begin{equation}
    \rho_{\rm DM}=0.45 \ \rm GeV/cm^3,
\end{equation}
from which we determine the axion amplitude $a_0$ as a function of $m_a$. 
In addition, $\alpha$ is a random phase of axion oscillation; it is approximately constant within the coherence time scale of DM during which the coherent oscillation (with constant $\alpha$) is maintained.  The coherence time of the DM is estimated as
\begin{equation}
    \tau_{\rm DM} \sim \frac{1}{m_a v^2_{\rm DM}} \sim 100 \ \mathrm{\mu s} \left(\frac{10 \ \mathrm{\mu eV}}{m_a}\right),
\end{equation}
where $v_{\rm DM}\sim 10^{-3}$ is the virial velocity of DM around the earth.

With the oscillating axion field, the electric field is induced if an external magnetic field is applied. Such a setup is often utilized in the haloscope searches of the axion DM \cite{ADMX:2018gho,CAPP:2020utb,HAYSTAC:2023cam,QUAX:2023gop}. In the following, we consider the induced electric field for the case with a background magnetic field $\vec{B_0}$ which is assumed to be static and homogeneous. We denote the vector potential describing the induced electric field as $A_\mu$, for which the following gauge conditions are imposed:
\begin{equation}
    \vec{\nabla} \cdot \vec{A}=0, \quad A_0=0.
\end{equation}
Expanding the electromagnetic field around the background magnetic field, the Lagrangian describing the dynamics of $A_\mu$ becomes
\begin{align}
    L = \int d^3 x \left[ 
    \frac{1}{2} \partial_\mu \vec{A} \cdot \partial^\mu \vec{A}
    + g_{a\gamma\gamma} \dot{a} \vec{A}  \cdot \vec{B}_0 
    \right],
    \label{L_agg}
\end{align}
where the ``dot'' on top of functions denotes the derivative with respect to time. The effect of the axion DM can be regarded as a small perturbation and, in Eq.\ \eqref{L_agg}, terms irrelevant to the following discussion are neglected. Note that electric field $\vec{E}\equiv -\dot{\vec{A}}$ and $\vec{A}$ are conjugated variables.

Because we consider qubits located inside a shielding cavity, which has a conductor wall, we should consider the electromagnetic field inside such a cavity. Particularly, the electric field should be perpendicular to the cavity wall at the surface of the wall. Hereafter, we quantize the gauge field $A_\mu$ and evaluate the electric field generated by the axion oscillation taking into account such a boundary condition.

We decompose the gauge field $\vec{A}$ and electric field $\vec{E}$ using the mode functions inside the cavity as
\begin{gather}
    \vec{A}(t, \vec{x})=-\sum_m \left(\frac{1}{2\omega_m }\right)^{1/2}(b_m+b^\dagger_m) \vec{E}_m(\vec{x}),\\
    \vec{E}(t, \vec{x})=i\sum_m \left(\frac{\omega_m}{2 }\right)^{1/2}(b_m-b^\dagger_m) \vec{E}_m(\vec{x}).
    \label{eq:Eexpress}
\end{gather}
The mode functions satisfy
\begin{subequations}\label{eq:modefunction}
    \begin{gather}
    (\vec{\nabla}^2 +\omega_m^2) \vec{E}_m=0,
    \label{Em1}\\    
    \vec{\nabla}\cdot \vec{E}_m=0,
    \label{Em2}\\
    \vec{E}_m |_{\parallel}=0,
    \label{Em3}
\end{gather}
\end{subequations}
where $\omega_m$ is the eigenfrequency of $m$-th cavity mode, and the subscript ``${\parallel}$'' denotes the projection of the vector onto the cavity wall at its surface. The mode functions are normalized as
\begin{align}
\int d^3 x \vec{E}_m (x) \cdot \vec{E}_n (x) = \delta_{mn},
\label{Em4}
\end{align}
where, in the above expression, the volume integral is performed inside the cavity. 
After the canonical quantization, $b_m$ and $b_m^\dagger$ satisfy 
\begin{align}
[b_m, b_n] = [b_m^\dagger, b_n^\dagger] =0, \quad
[b_m, b_n^\dagger] = \delta_{mn},
\end{align}
and hence they can be regarded as annihilation and creation operators. The Hamiltonian reads as
\begin{equation}
    H_0=\sum_m 
    \left[
    \omega_m b_m^\dagger b_m -2 g_m (b_m+b^\dagger_m)\sin (m_at-\alpha)
    \right],
    \label{eq:H_Ea}
\end{equation}
where
\begin{gather}
    g_m \equiv\frac{1}{2} g_{a\gamma \gamma} a_0 \left( \frac{1}{2 \omega_m} \right)^{1/2}  m_a  \int \vec{E}_m(\vec{x})  \cdot \vec{B}_0 \ d^3x. \label{eq:gm}
\end{gather}

We can derive the evolution equation of $b_m$ in the Heisenberg picture as
\begin{align}
\frac{d}{dt} b_m (t) = -i[b_m (t),H(t)] 
=-i\omega_m b_m(t) + g_m (e^{i (m_at -\alpha)}- e^{-i(m_at-\alpha)}),
\end{align}
whose special solution is given by
\begin{equation}
    b_m(t)= g_m \left( \frac{e^{i(m_at -\alpha)}}{i(\omega_m+m_a)}  - \frac{e^{-i(m_a t- \alpha)}}{i(\omega_m-m_a)} \right).
\end{equation}
We neglect the term proportional to $e^{-i\omega_m t}$ because such a term should dissipate away. The expectation value of the electric field becomes
\begin{gather}
    \vec{E}(t,\vec{x})= \cos (m_a t-\alpha) \sum_m\frac{m_a^2}{m_a^2-\omega_m^2} g_{a\gamma \gamma }a_0 \vec{E}_m(\vec{x})  \int d^3x' \vec{E}_m (\vec{x}') \cdot \vec{B}_0. \label{eq:Esol}
\end{gather}
Thus, with the background magnetic field, the axion DM generates an AC electric field. Notice that the above result agrees with that obtained by solving the classical Maxwell equation coupled to the oscillating axion (for details, see Appendix~\ref{appendix:classic}).

\section{Transmon excitation} 
\label{sec:transmonex}
\setcounter{equation}{0}

In this section, we discuss the excitation of a transmon qubit owing to the presence of an axion-induced electric field, taking into account the effect of the shielding cavity. The transmon qubit consists of a capacitor and a Josephson junction; for the axion detection with the transmon qubit, we use the interaction of the electric field with the capacitor component. As we have seen, the axion oscillation induces an AC electric field once an external magnetic field is applied. With the oscillating axion field of the form given in Eq.\ \eqref{a(t)}, the induced electric field is given in the following form:
\begin{align}
    \vec{E} = \vec{\bar{E}} \cos (m_a t-\alpha),
\end{align}
with $|\vec{\bar{E}}|\equiv \bar{E}$. We parameterize the size of the axion-induced electric field as
\begin{equation}
    \bar{E} \equiv g_{a\gamma\gamma} a_0 B_0 \kappa, \label{eq:Ebar}
\end{equation}
where $\kappa$ is a numerical constant; $\kappa=1$ if we consider the case in the vacuum while $|\kappa|>1$ is possible inside the cavity. Based on the argument in the previous section, which neglects the mixing of qubit and cavity modes, $\kappa$ is estimated as
\begin{gather}
    \kappa = \sum_m \vec{E}_m (\vec{x}) \cdot  \hat{z} \left[ \frac{m_a^2}{m_a^2-\omega_m^2}\int dV \vec{E}_m \cdot \vec{B}_0/B_0 \right],
    \label{eq:kappa}
\end{gather}
where the direction of $\vec{E}$ is chosen to be the $z$ direction and $\hat{z}$ is the unit vector pointing to the $z$ direction. 
Eq.\ \eqref{eq:kappa} indicates that the $\kappa$ parameter is enhanced when $m_a$ is close to one of the cavity frequencies. 

The detailed derivation of the effective Hamiltonian of the transmon qubit coupled to the electric field can be found in Ref.\ \cite{Chen:2022quj}, which is utilized in the following analysis. Under the presence of an axion-induced electric field, the effective Hamiltonian of transmon is given as
\begin{equation}
    H=\omega_{\rm q} \ket{e}\bra{e} + 2\eta \cos(m_a t-\alpha)(\ket{e}\bra{g}+\ket{g}\bra{e}),
    \label{eq:Hamil}
\end{equation}
where $\ket{g}$ and $\ket{e}$ are the ground and excited states of the transmon qubit, respectively. The coupling strength $\eta$ is given by
\begin{equation}
    \eta \equiv \frac{\sqrt{\omega_{\rm q} C}}{2 \sqrt{2}} \bar{E} d
    \cos \Theta,    
    \label{eq:eta}
\end{equation}
where $C$ and $d$ are the capacitance and the distance between the capacitor plates of the transmon, respectively (see, e.g., Ref.\ \cite{S_miller2018} for the derivation of distance $d$). The parameter $\Theta$ is the angle between $\vec{\bar{E}}$ and the normal vector of the capacitor; hereafter, we consider the case of $\Theta=0$. Note that the second term in Eq.~\eqref{eq:Hamil} tells us that the axion-induced electric field can stimulate the state-transition of the transmon qubit.
We should be aware that when $\omega_{\rm q}$ is close to one of the cavity frequencies $\omega_{m=c}$ (in particular, when $|\omega_{\rm q}-\omega_{c}|\lesssim \lambda_{c}$, where $\lambda_c$ is the coupling between the cavity mode and qubit which is of order $\sim O(1)$ MHz), one needs to take into account the quantum mode mixing between the qubit state and the cavity mode. We deal with such a particular case in Appendix \ref{appendix:strongcoupling}.

The evolution of the qubit state, denoted as $\ket{\Psi}$, is governed by the following Schrodinger equation:
\begin{align}
    i \frac{d}{dt} \ket{\Psi (t)} = H \ket{\Psi (t)}. 
\end{align}
Denoting the (time-independent) ground and excited states of the qubit as $\ket{g}$ and $\ket{e}$, respectively, we decompose the qubit state into the following form:
\begin{equation}
    \ket{\Psi (t)}=\psi_g(t)\ket{g}+\psi_e(t) e^{-i\omega_{\rm q} t}\ket{e}.
\end{equation}
Then, the wave functions obey the following differential equation:
\begin{align}
    i 
    \begin{pmatrix}
        \dot{\psi}_g \\ \dot{\psi_e}
    \end{pmatrix}
    = 2 \eta \cos (m_a t - \alpha) 
    \begin{pmatrix}
        e^{-i\omega_{\rm q}t} \psi_e \\ e^{i\omega_{\rm q}t} \psi_g
    \end{pmatrix}.
\end{align}

Within the coherence time $\tau$ of the system, the qubit undergoes a unitary evolution governed by the Hamiltonian given in Eq.\ \eqref{eq:Hamil}; the coherence time $\tau$ is approximately given by ${\rm min}(\tau_q, \tau_{\rm DM})$, where $\tau_q$ and $\tau_{\rm DM}$ are the coherence time of the qubit and that of the DM, respectively. We can define a unitary matrix (or operator) $U_{\rm DM}(t)$ describing the unitary evolution, with which we obtain
\begin{equation}
    \begin{pmatrix}
    \psi_g(t)\\
    \psi_e(t)
    \end{pmatrix}
    = 
    U_{\rm DM} (t)
    \begin{pmatrix}
    \psi_g(0)\\
    \psi_e(0)
    \end{pmatrix}.
\end{equation}
Because the effect of the DM axion is regarded as a small perturbation to the evolution of the qubit, it is instructive to expand $U_{\rm DM}$ in terms of $\eta$ as
\begin{align}
    U_{\rm DM}(t) \simeq
    \begin{pmatrix}
        1 & -\eta ( 
        e^{-i\alpha}\frac{e^{i(m_a-\omega_{\rm q})t}-1}{m_a-\omega_{\rm q}}
        - e^{i\alpha}\frac{e^{-i(m_a+\omega_{\rm q})t}-1}{m_a+\omega_{\rm q}})
        \\
        \eta ( 
        e^{i\alpha}\frac{e^{-i(m_a-\omega_{\rm q})t}-1}{m_a-\omega_{\rm q}}
        - e^{-i\alpha}\frac{e^{i(m_a+\omega_{\rm q})t}-1}{m_a+\omega_{\rm q}})
        & 1
    \end{pmatrix},
    \label{U_DM}
\end{align}
where terms of order $\eta^2$ are neglected.  We can see that the excitation rate of the qubit is enhanced when the qubit frequency $\omega_{\rm q}$ is close to the oscillation frequency of the axion DM. Indeed, for a long enough time period satisfying $(m_a+\omega_{\rm q})^{-1}\ll t\lesssim |m_a-\omega_{\rm q}|^{-1}$, $U_{\rm DM}(t)$ is approximated as
\begin{align}
    U_{\rm DM} (t)\xrightarrow{(m_a+\omega_{\rm q})^{-1}\ll t\lesssim |m_a-\omega_{\rm q}|^{-1}}
    \begin{pmatrix}
        1 & -i e^{-i\alpha} \eta t \\
        -i e^{i\alpha} \eta t & 1
    \end{pmatrix},
    \label{UDM(smalltau)}
\end{align}
so that, if the qubit is initialized to the ground state, the transition probability from $\ket{g}$ to $\ket{e}$ is given by $\eta^2 t^2$. For DM detection, a longer evolution time period $t$ is more advantageous because the transition probability is proportional to $t^2$. 

In our protocol of the axion DM search with transmon qubit, we propose to take the evolution time period of the qubit to be as long as the coherence time of the system $\tau$; such a choice of the evolution time period maximizes the sensitivity.  Such a detection experiment (with fixed $\omega_{\rm q}$) is sensitive to the DM with its mass in the range of $\omega_{\rm q}-\tau^{-1}\lesssim m_a\lesssim\omega_{\rm q}+\tau^{-1}$; for the axion DM within such a mass range, the transition probability of the qubit for one measurement cycle is $\sim\eta^2\tau^2$. In performing the frequency scan, we assume to take the step width to be $\tau^{-1}$.  

Hereafter, we consider the resonance limit in which the axion mass is equal to the qubit frequency, which well approximates the qubit evolution for the case of $\omega_{\rm q}-\tau^{-1}\lesssim m_a\lesssim\omega_{\rm q}+\tau^{-1}$.  Adopting the rotational-wave approximation, the evolution unitary matrix is given by
\begin{equation}
U_{\rm DM} \simeq
\begin{pmatrix}
\cos \eta \tau & -ie^{-i\alpha} \sin \eta \tau\\
-ie^{i\alpha} \sin \eta \tau & \cos \eta \tau
\end{pmatrix}.
    \label{eq:qubitevolve}
\end{equation}
Here and hereafter, the evolution time is taken to be $\tau$ so that $U_{\rm DM}$ should be understood as $U_{\rm DM}(\tau)$. Notice that $U_{\rm DM}$ given in Eqs.\ \eqref{UDM(smalltau)} and \eqref{eq:qubitevolve} agrees when $\tau\ll\eta^{-1}$. The unitary operator $U_{\rm DM}$ has the following eigenstates:
\begin{equation}
    \ket{\psi_{\pm}}=
    \frac{1}{\sqrt{2}}
    (\ket{g}\mp e^{i\alpha} \ket{e}) \label{eq:eigenstate}
\end{equation}
with the corresponding eigenvalue $e^{\pm i\eta \tau}$. Notice that for convenience we absorbed the evolution phase $e^{-i\omega_{\rm q} \tau}$ from the free part of Hamiltonian into the definition of $\ket{e}$ state, which we shall apply also hereafter.
In addition, with qubit initialized to the ground state as $\psi_g(0)=1$ and $\psi_e(0)=0$, we obtain the probability to observe a qubit to be at the excited state at $t=\tau$ as
\begin{align}
    p_{g\rightarrow e}^{(1)} \simeq \eta^2 \tau^2,
    \label{eq:pge}
\end{align}
assuming that $\eta \tau \ll 1$. Here, the superscript ``(1)'' indicates the transition probability for one qubit. Numerically, we find
\begin{align}
    p_{g\rightarrow e}^{(1)} \simeq 0.11  \times &\left( \frac{g_{a\gamma \gamma}}{10^{-10} \ \mathrm{GeV^{-1}}}\right)^2  \left( \frac{m_a}{1 \ \mathrm{\mu eV}}\right)^{-1}
    \left( \frac{B_0}{1 \ \mathrm{T}}\right)^2\left( \frac{\tau}{100 \ \mathrm{\mu s}} \right)^2 \kappa^2 \nonumber \\
    &\times \left( \frac{C}{0.1 \ {\rm pF}}\right)
    \left( \frac{d}{100 \ {\rm \mu m}}\right)^2
    \left( \frac{\rho_{\rm DM}}{0.45 \ {\rm GeV/cm^3}}\right).
\end{align}

\section{Protocol and sensitivity} 
\label{sec:protocol&sen}
\setcounter{equation}{0}

In the following, we elaborate on the steps to perform a DM detection experiment utilizing transmon qubit and show the sensitivity we expect. We consider two protocols for axion DM detection. After summarizing the experimental setup of our assumption in subsection \ref{subsec:setup}, we first consider a protocol in which a number of qubits are treated individually, in which one simply prepares a number of qubits and looks for the excitation of individual qubits (subsection \ref{subsec:separabel}). Then, in subsection \ref{subsec:GHZ}, we discuss a protocol to use entangled qubits to enhance the signal rate. We introduce a quantum circuit to accumulate information from DM coherently, resulting in the improved scaling of the sensitivity to the axion-photon-photon coupling parameter with respect to the number of qubits.

\subsection{Experimental setup}
\label{subsec:setup}

To make our discussion concrete, we assume a simple experimental setup for our estimation. We assume that the qubits are located in a cylindrically-shaped shielding cavity, which has the radius $R$ with its axis pointing in the $z$ direction. We also assume the uniform external magnetic field $\vec{B}_0$ parallel to the cylinder axis. 

In such a setup, the axion-induced electric field is pointing to the $z$-direction. Then, because $\vec{\nabla} \cdot \vec{E}=0$, $\vec{E}$ is $z$-independent. Based on this observation, the induced electric field can be expanded by the mode functions of the following form:
\begin{equation}
\vec{E}_m(\vec{x})=\hat{z}\frac{J_0(\omega_m r)}{\sqrt{V} J_1(\omega_m R)},
\end{equation}
where $r$ is the radial coordinate for the plane perpendicular to the $z$-axis (i.e., the distance from the cylinder axis), $V$ is the volume of the shielding cavity, and $J_{0}$ and $J_1$ are the Bessel functions of the first kind. In addition, $\omega_m$ denotes the cavity frequency, which is given by $\omega_m=x_m/R$ with $x_m$ being the $m$-th zero of $J_0$. By using identities
\begin{equation}
    \int^{x_n}_0 x J_0(x) J_0\left(x \frac{x'}{x_n} \right) dx= \frac{x_n^2}{x_n^2-x'^2} x_n J_0(x')J_1(x_n),
\end{equation}
and
\begin{equation}
    \int^{x_n}_0 x J_0^2(x)dx=\frac{1}{2}x_n^2 J_1^2(x_n),
\end{equation}
as well as the Fourier-Bessel series, the factor $\kappa$ given in Eq.~\eqref{eq:kappa} can be simplified to
\begin{equation}
    \kappa=1-\frac{J_0(m_a r)}{J_0(m_a R)}.
    \label{kappa_cylinder}
\end{equation}
Notice that, as discussed in the previous section, the above expression is not applicable when $\omega_{\rm q}$ is close to one of the cavity-mode frequencies.

Inside the cavity, we assume that the qubits are located at the center of the cylinder (i.e., $r=0$).  For the estimation of the discovery reach for the axion DM using transmon qubits, the circuit parameters of the qubits are taken as follows:
\begin{itemize}
\item $C=0.1$ pF.
\item $d=100\ \rm \mu m$.
\end{itemize}    
We also assume $B_0=5\ {\rm T}$.  In addition, the number of the qubits is denoted as $n_{\rm q}$; we consider the cases of $n_{\rm q}=1$ and $100$.

For the qubit frequency tunability, we assume a type of transmons equipped with a superconducting quantum interference device (SQUID), i.e., the qubit frequency varies according to the change of small ($O(0.1)\, {\rm m T}$) external magnetic flux applied through the SQUID loop.  Notice that the magnetic field necessary to tune the qubit frequency is perpendicular to the transmon thin film as well as the strong magnetic field for the axion conversion. Thus, in principle, these two magnetic fields do not interfere and are independent.

\subsection{Protocol with individual measurement}
\label{subsec:separabel}

Now we are in the position to examine the expected sensitivity of the axion DM search using the transmon qubits. First, we discuss a simple detection method in which each qubit is regarded as an independent quantum sensor and is read out individually. 

The following is the protocol we propose to detect the DM signal using separate preparation and measurement of a number of qubits:
\begin{enumerate}
    \item Prepare $n_{\rm q}$ qubits.
    \item Initialize them to their ground states and wait as long as the coherence time of the system $\tau$. Then, measure the state of each qubit.
    \item Repeat the process $N_{\rm try}$ times until the desired interrogation time.
    \item Change the qubit frequency to scan the axion mass and repeat the whole protocol.
\end{enumerate}

For each fixed qubit frequency, the number of the excited qubit is given by
\begin{equation}
    N_{\rm sig}=N_{\rm try} n_{\rm q} p_{g\rightarrow e}^{(1)}
    \simeq N_{\rm try} n_{\rm q} \eta^2 \tau^2,
    \label{eq:linearscale}
\end{equation}
which corresponds to the total number of signals at a single frequency bin. On the other hand, the expected background events are due to the readout error. We take into account the effect of unexpected noises during the qubit evolution by a properly chosen value of coherence time $\tau$. Based on the current qubit technology achieving single-shot ground-state readout fidelity of $\sim99.9$ \% \cite{Chen:2022frn}
as well as on expectations of the near-term physical error rate satisfying the error correction threshold for realizing the fault-tolerant quantum computer \cite{Fowler:2012hwn,Bravyi:2023qpn},
we assume that the readout error rate is $p_{\rm r}= 0.1 \ \%$ in our calculation. The number of background events for each frequency bin is then given by
\begin{equation}
    N_{\rm bkg}= N_{\rm try} n_{\rm q} p_{\rm r}.
\end{equation}
We define the significance $\sigma$ by a signal over the fluctuation of the background:
\begin{equation}
    \sigma =\frac{N_{\rm sig}}{\sqrt{N_{\rm bkg}}}= \frac{N_{\rm try}n_{\rm q} \eta^2 \tau^2}{\sqrt{N_{\rm try}  n_{\rm q}p_{\rm r} }}.
    \label{eq:sig1}
\end{equation}
Notice that, because we expect to perform the scan over the qubit frequencies, we may use the side-band method~\cite{Aad2012} to understand the background; the background level can be estimated from the study of the case with the qubit frequency $\omega_{\rm q}$ not matched with axion mass (i.e., side-band) because the qubit excitation due to the axion DM is expected to be sizable only when $\omega_{\rm q}$ is close to $m_a$ (see Eq.\ \eqref{U_DM}). Our criterion of the discovery reach of the axion DM is based on $\sigma \geq 5$. 

\begin{figure}[t]
  \centering
  \includegraphics[width=0.7 \linewidth]{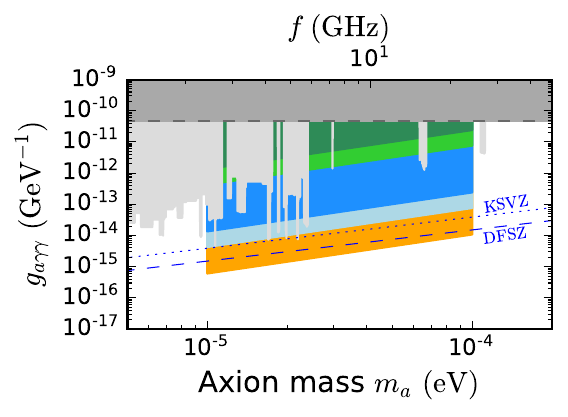}
  \caption{Projection plot regarding the $5\sigma$ sensitivity of axion-photon coupling $g_{a\gamma\gamma}$. We assume a year experiment with the coherence time of the system determined by that of DM ($\tau=10^6/m_a$). We assume static magnetic field $B_0=5$ T applied uniformly pointed along the cylinder axis of the shielding cavity. The dark and light green (blue) contours correspond to the reach with $\kappa=1 \ (100)$ using $n_{\rm q}=1$ and $n_{\rm q}=100$, respectively, where the individual measurement protocol is applied. On the other hand, the orange contour corresponds to the reach from using entangled qubit sensors with $n_{\rm q}=100$ and $\kappa=100$ assumed. The readout error rate and gate operation error rate are both assumed to be 0.1\%.  The dark grey area shows the astrophysical constraint from the observation of the stellar population in the Globular Clusters \cite{Dolan:2022kul}, while the grey area shows the constraints from Haloscope experiments \cite{AxionLimits}. The dotted and dashed blue lines are parameter regions suggested by the KSVZ model and DFSZ model of QCD axions, respectively.}
  \label{fig:k1cavityreac}
\end{figure}

We first show the expected sensitivity to the $g_{a\gamma\gamma}$ parameter with fixing the value of $\kappa$ parameter. The discovery reaches with $\kappa=1$ and $100$ are shown in Fig.\ \ref{fig:k1cavityreac} as functions of the axion mass. (In such cases, the strong mode mixing between qubit and cavity is avoided. This can be checked by $|\omega_{\rm q}-\omega_c| \simeq |m_a-\omega_c| \simeq   \frac{m_a}{ \kappa} \gg \lambda_c$ where $\omega_c$ is the dominant cavity mode relevant.) We consider the case to scan the axion mass range of $10 \ \mathrm{\mu eV}<m_a<100\ {\rm \mu eV}$ within 1 year. In our numerical analysis, the coherence time $\tau$ of the system is taken to be equal to $\tau_{\rm DM}=10^6/m_a$, assuming that the coherence time of the qubit and cavity are longer than $\tau_{\rm DM}$. This is an achievable assumption since qubit coherence time longer than $100 \ {\rm\mu s}$ is already commonly achieved in the experiments \cite{Kjaergaard:2019lmy,Place:2021elq,Wang_2022}. We can see that, even with $n_{\rm q}\sim 1$ and $\kappa\sim 1$, we may reach the parameter region which has not been excluded yet. With a larger value of $n_{\rm q}$ or $\kappa$, sensitivity becomes better as expected. In particular, with $n_{\rm q}\sim 100$ and $\kappa\sim 100$, we may reach the parameter region suggested by a QCD axion model. Note that $|\kappa| \gg 1$ condition can be realized and maintained by adaptively tuning the frequency of the cavity. Also, notice that, when $R\gg 1/m_a$, $|\kappa|$ approximately scales as $\sim\sqrt{m_a R}$ (see Eq.~\eqref{kappa_cylinder}) unless $m_a$ is close to any of cavity frequencies.

\begin{figure}[t]
  \centering
  \includegraphics[width=0.7 \linewidth]{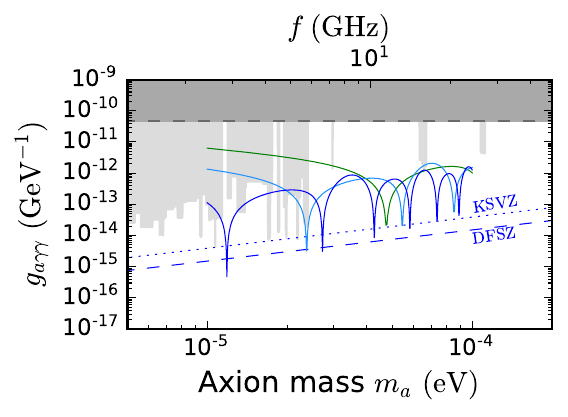}
  \caption{Projection plot regarding the $5\sigma$ sensitivity of axion-photon coupling $g_{
a\gamma\gamma}$. We assume a year experiment with the coherence time of the system determined by that of DM ($\tau=10^6/m_a$). We consider, as an example, a cavity with fixed radius $R=1,2,4$ cm (for green, light blue, and blue lines) and height $2R$ with magnetic field $B_0=5$ T applied uniformly pointed along the cylinder axis of the shielding cavity.
The readout error rate is assumed to be 0.1\%. We assume $n_{\rm q}=100$ number of qubits are used.
The dark grey area shows the astrophysical constraint from the observation of the stellar population in the Globular Clusters \cite{Dolan:2022kul}, while the grey area shows the constraints from Haloscope experiments \cite{AxionLimits}. The dotted and dashed blue lines are parameter regions suggested by the KSVZ model and DFSZ model of QCD axions, respectively.}
  \label{fig:fixcavityreac}
\end{figure}

It is also instructive to consider the case with a fixed-size cavity.  In Fig.~\ref{fig:fixcavityreac}, we show the expected discovery reach, assuming cavities of $R=1$, $2$, and $4\ {\rm cm}$.  Again, we assume to scan the range of $10 \ {\rm \mu eV} < m_a <100\ {\rm \mu eV}$ within 1 year. In this case, we can find significant enhancements in the sensitivity for particular values of the axion mass.  Such enhancements occur when $m_a$ is close to one of the cavity frequencies; in such a case, the $\kappa$ parameter is enhanced due to the excitation of the electric field whose frequency is close to the axion mass. Note that at the very peak positions of the contours when the tuned frequency of qubit $\omega_{\rm q}$ is closely matched to one of the cavity frequencies, one needs to take into account the hybridized mode excitation of the qubit and cavity, of which we perform the treatment and elaborate on the detail in Appendix \ref{appendix:strongcoupling}. As a result, with the cavity effect, we may be able to probe various types of QCD axion models (i.e., KSVZ and DFSZ models) with using transmon qubits as quantum sensors.

\subsection{Detection  protocol with entangled qubits}
\label{subsec:GHZ}

The sensitivity may be drastically improved if we use a multiple of entangled qubits. Such a possibility has been pointed out in Ref.\ \cite{Chen:2023swh} of which we briefly explain the idea here. We concentrate on the case without the strong mode mixing between the shielding cavity and qubits.  We also assume that the coherence time of individual qubits is so long that the coherence time of the system is determined by DM and not by the qubits.

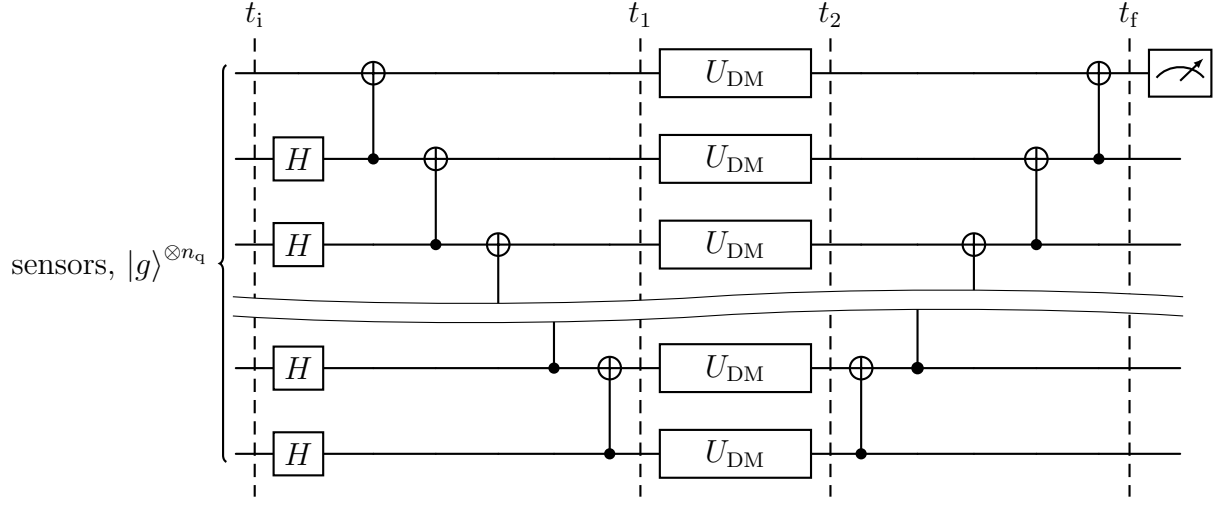
\begin{figure*}[t]
  \begin{quantikz}
    \lstick[6]{sensors, $\ket{g}^{\otimes n_{\rm q}}$} \slice[style=black]{$t_{\rm i}$} && \targ{}  &  & && \slice[style=black]{$t_{\rm 1}$} & \gate[1][2 cm]{U_{\rm DM}} \slice[style=black]{$t_{\rm 2}$} & && &   &  \targ{ } \slice[style=black]{$t_{\rm f}$} &\meter{} 
    \\
     &\gate{H}& \ctrl{-1} & \targ{}& & &&\gate[1][2 cm]{U_{\rm DM}} & & & & \targ{} & \ctrl{-1}&
    \\
      &\gate{H}& &\ctrl{-1} & \targ{}\vqw{1} & &&\gate[1][2 cm]{U_{\rm DM}} & & &  \targ{}\vqw{1} & \ctrl{-1} &&
    \\
    \wave&&&&&&&&&&&&&
    \\
      &\gate{H}& &&& \ctrl{-1} & \targ{} & \gate[1][2 cm]{U_{\rm DM}}&\targ{}& \ctrl{-1} \ & & & & 
     \\
      &\gate{H}& && && \ctrl{-1} & \gate[1][2 cm]{U_{\rm DM}} &\ctrl{-1} & & & & &
\end{quantikz}
  \caption{Quantum circuit for the DM detection. The gate with $H$ represents the Hadamard gate, while that with ``$\bullet$'' and ``$\oplus$'' connected by the line is the CNOT gate (where ``$\bullet$'' is the control qubit). The $U_{\rm DM}$ represents the evolution with the effect of DM. Figure from Ref~\cite{Chen:2023swh}.}
  \label{fig:circuit}
\end{figure*}

In Fig.\ \ref{fig:circuit}, for readers' convenience, we show the quantum circuit which explains the procedure of detecting axion DM using entangled qubits, proposed in Ref.\ \cite{Chen:2023swh}. We assume that $t_1-t_{\rm i} \sim t_{\rm f}-t_2 \ll t_2-t_1$, so that we have to consider the effect of DM on the evolution of the qubit system only when $t_1 \leq t \leq t_2$;  $t_2-t_1$ will be set to be around the coherence time of the system $\tau$. To see the enhancement mechanism, let us illustrate the case when the phase of the DM field is given by $\alpha=0$. Initially, the qubit state $\ket{\Psi}$ is prepared as 
\begin{align}
  \ket{ \Psi (t_i) } = \ket{g}^{\otimes n_{\rm q}}.
\end{align}
Then, after performing the CNOT gates, the qubit states at $t=t_1$ is given by
\begin{align}
  \ket{ \Psi (t_1) } = 
  \frac{1}{\sqrt{2}}
  \left( 
  \ket{+}^{\otimes n_{\rm q}} +
  \ket{-}^{\otimes n_{\rm q}}
  \right),
  \label{psi_init}
\end{align}
where 
\begin{align}
  \ket{\pm} \equiv \frac{1}{2} (\ket{g} \pm \ket{e} ).
\end{align}
Notice that $\ket{ \Psi (t_1) }$ is nothing but the Greenberger–Horne–Zeilinger (GHZ) state made of $n_{\rm q}$ qubits \cite{10.1119/1.16243}.  Since $\alpha=0$, $\ket{+}$ and $\ket{-}$ are the eigenstates of the unitary operator $U_{\rm DM}$ (see Eq.\ \eqref{eq:qubitevolve}). Then, the qubit state after the evolution with the effect of the DM is given by
\begin{align}
  \ket{\Psi (t_2)} = 
  &\frac{1}{\sqrt{2}}
  \left( 
  e^{-in_{\rm q}\eta \tau} \ket{+}^{\otimes n_{\rm q}} +
  e^{+in_{\rm q} \eta \tau} \ket{-}^{\otimes n_{\rm q}}
  \right).
 \label{eq:psi_t2}
\end{align}
We can see that the $\ket{+}^{\otimes n_{\rm q}}$ and $\ket{-}^{\otimes n_{\rm q}}$ acquire different signs of phases while the phases are coherently accumulated and are proportional to the number of qubits. After operating the CNOT gates again, the phase information is moved to the first qubit as:
\begin{align}
  \ket{\Psi (t_{\rm f})} &= \frac{1}{\sqrt{2}}
  \left( e^{-in_{\rm q} \eta \tau}\ket{+}+ e^{+in_{\rm q} \eta \tau}\ket{-} \right)\otimes \ket{+}^{\otimes (n_{\rm q}-1)}
  \nonumber \\
  &= \left[ \cos(n_{\rm q} \eta \tau) \ket{g}- i\sin(n_{\rm q} \eta \tau) \ket{e} \right] \otimes \ket{+}^{\otimes (n_{\rm q}-1)},
 \label{eq:psi_tf}
\end{align}
giving the excitation probability of the first qubit as $\simeq n_{\rm q}^2 \eta^2 \tau^2$. Similarly, in the case with a general value of $\alpha$, it can be shown that the excitation probability with the GHZ state becomes
\begin{align}
    p^{\rm (GHZ)}_{g\rightarrow e} \simeq
    n_{\rm q}^2\eta^2 \tau^2 \cos^2 \alpha
    + O(n_{\rm q})
    \simeq \frac{1}{2} n_{\rm q}^2\eta^2 \tau^2,
\end{align}
where, in the second equality, we have used the fact that the phase $\alpha$ is expected to take a random value for each measurement cycle (as far as $t_f-t_i\gtrsim\tau_{\rm DM}$). The excitation of the first qubit is regarded as the signal of DM. Then, the number of events for the case with the GHZ state is estimated as
\begin{align}
    N_{\rm sig}^{\rm (GHZ)} \simeq 
    N_{\rm try} p_{g\rightarrow e}^{\rm (GHZ)}
    \simeq \frac{1}{2} N_{\rm try} n_{\rm q}^2
    \eta^2 \tau^2.
    \label{S(GHZ)}
\end{align}
Comparing Eqs.\ \eqref{eq:linearscale} and \eqref{S(GHZ)}, we can see a significant enhancement of the number of signals for the case with the GHZ state when $n_{\rm q}\gg 1$; the number of signal scales as $N_{\rm sig}\propto n_{\rm q}$ for the case of the individual measurement while it scales as $N_{\rm sig}^{\rm (GHZ)}\propto n_{\rm q}^2$ for the case with the GHZ state.

Notice that, with our quantum circuit, we need $O(n_{\rm q})$ gate operations for each measurement cycle, so we need to take into account the false excitation rate coming from the gate operator error. For the DM detection with GHZ state, an improvement in the reliability of the gate operation is desirable; we expect that such an improvement can be realized in the future when quantum computers with a large number of qubits become available. Because, currently, we cannot estimate the error rate of the gate operations in the future, here we make a simple estimation of the error rate. Assuming that the error rate of each gate operation is $\sim p_{\rm op}$, we parameterize the false excitation rate per one measurement cycle as $p_{\rm r} + n_{\rm q}p_{\rm op}$, which gives the number of background for $N_{\rm try}$ repetitions of measurement cycles as
\begin{equation}
    N_{\rm bkg}^{\rm (GHZ)}
    =N_{\rm try} (p_{\rm r} + n_{\rm q}p_{\rm op}).
\end{equation}
Numerically, we take $p_{\rm op}=0.1\ \%$ in our estimation, based on the fact that with the state-of-the-art technology \cite{Li:2023cla,Li:2024xbj} the fidelity above 99.9 \% is already achieved for both two-qubit gate and single-qubit gate operations. 

The protocol to detect the DM axion with GHZ state is similar to the case of individual measurement discussed in the previous subsection. The protocol is given as follows:
\begin{enumerate}
    \item Prepare $n_{\rm q}$ qubits.
    \item Initialize them to the ground state and pass them through the quantum circuit shown in Fig.\ \ref{fig:circuit}. The time interval $t_2-t_1$ should be as long as the coherence time $\tau$ (assuming that the coherence of the entangled system can be maintained during such a time interval). Then, perform the measurement of the readout qubit (i.e., the first qubit).
    \item Repeat the measurement cycle $N_{\rm try}$ times until the interrogation time.
    \item Change the qubit frequency to scan the axion mass and repeat the whole protocol.
\end{enumerate}
We note that the entangled qubit states are more fragile than the quantum state made of a single qubit; the coherence time of the entangled system is roughly given by $\sim \tau_{\rm q}/n_{\rm q}$, where $\tau_{\rm q}$ is the coherence time of a single qubit. Therefore, to maintain the coherence during $\tau$, the qubit system needs to have a longer coherence time than that of DM; based on this constraint, the number of qubits for this protocol is bounded from above as $n_{\rm q} \lesssim \tau_{\rm q}/ \tau_{\rm DM}$.

We also estimate the discovery reach of the axion DM for the case with the entangled GHZ state, taking $n_{\rm q}=100$.  Here, we adopt the significance of 
\begin{equation}
    \sigma^{\rm (GHZ)} = 
    \frac{N_{\rm try}n^2_{\rm q} \eta^2 \tau^2/2}{\sqrt{N_{\rm try}  ( p_{\rm r} + n_{\rm q}p_{\rm op} ) }}.
    \label{sigma(GHZ)}
\end{equation}
The result is also shown in Fig.\ \ref{fig:k1cavityreac}, in which we can observe a significant improvement in the reach. In particular, we may have a chance to access various types of QCD axion models (like the KSVZ and DFSZ models).  We also note here that, comparing Eqs.\ \eqref{eq:sig1} and \eqref{sigma(GHZ)}, $\sigma\propto n_{\rm q}^{1/2}$ while $\sigma^{\rm (GHZ)}\propto n_{\rm q}^{3/2}$ in the large $n_{\rm q}$ limit. 


\section{Conclusions and discussion} 
\label{sec:d&c}
\setcounter{equation}{0}

We have presented the possibility of using superconducting transmon qubits to directly detect axion DM. Since transmon interacts with an electromagnetic field through its capacitor, it has the sensitivity to the electric field induced by axion DM under the static magnetic field applied.
With a simple 1-year experiment measuring the qubit excitation separately owing to the effect of DM, we have shown that conservatively the axion-photon coupling can be probed down to $g_{a\gamma\gamma}\simeq 10^{-12} \ (10^{-13})  \ \rm{GeV^{-1}}$ using $n_{\rm q}=1, \ (100)$ qubits. We have also pointed out drastic improvements of the sensitivity with the use of (i) the cavity effect to amplify the induced electric field and (ii) the quantum circuit to accumulate coherently the information of axion DM proposed in Ref.~\cite{Chen:2023swh}. Regarding the cavity effect, when the mass of axion $m_a$ becomes close to one of the cavity mode frequencies, the near-resonant excitation of the electric field occurs, resulting in the amplification of the signal. On the other hand, on the enhancement using the entangled state, the signal rate scales not linearly but quadratically with the number of qubits in use, improving the sensitivity significantly. By incorporating the cavity effect and the entangled state, one can reach the QCD axion model including that of KSVZ and DFSZ models using, e.g., $n_{\rm q}= 100$. 

The challenges to overcome are several. One is to maintain the long coherent operation of qubits under the strong magnetic field which is the element needed for the conversion of axions to photons. In our setup with transmon thin film, the strong magnetic field necessary for the axion conversion is in the direction parallel to the thin film such that the electric field induced by axions can couple efficiently to the capacitor of the transmon; at the same time, this is the setup in which the transmon qubit is least sensitive to the magnetic field. At present, the magnetic field strength up to $O(1)$ T is shown to be applicable in this setup while maintaining the coherent operation of transmon. Alternatively, this problem may be dealt with by setting up the experiment separating the space applied by the strong magnetic field and the space containing transmon qubits \cite{Kono:2018rhn}. 
Other points are the ability to sweep conveniently the qubit frequency in the range of interest, and to arrange all of the qubit's frequencies to be of the same values within the bandwidth of the DM to realize the quantum enhancement properly when one uses a quantum circuit and GHZ state. 

Apart from the challenges mentioned, to improve further the sensitivity of the search, the longer coherence time of the qubit and a lower error rate of the readout and gate operations are desirable. The longer coherence time of qubits helps lengthen the time for the accumulation of the DM signal coherently; also, regarding the use of quantum circuits, the GHZ state is less fragile which allows us to use more number of entangled qubits. On the other hand, the lower error rate of the readout and gate operations helps improve the signal-to-noise ratio of the search.

With the rapid growth of the quantum sensor and quantum computing technology including the advancement of the noisy intermediate-scale quantum computer \cite{IBM,ionQ,Arute:2019zxq}, the useful characteristics of qubits such as an efficient tunability of the qubit, long coherence time and low error rate are already achieved to a certain extent and expected to be drastically improved further during the time.  In addition, when the large-scale production of qubits and fault-tolerant quantum computers becomes available, they shall at the next level bring a radical improvement in the sensitivity of DM detection using the qubits and quantum circuits of our proposal with an open end.
This also enables the potential of performing the DM search experiment using the current or future quantum computer platform.

\vspace{2mm}
\noindent
\underline{\it Acknowledgments}:
S.C. was supported by JSPS KAKENHI Grants No.\ 23K13093. T.I. was supported by JSPS KAKENHI Grants No.\ 23H04864 and JST PRESTO Grant No.\ JPMJPR2253, Japan. 
H.F. was supported by JSPS KAKENHI Grant No.\ 24K17042.
T.M. was supported by JSPS KAKENHI Grant No.\ 23K22486.
T.N. was supported by JSPS KAKENHI Grants No. 23K17688, No. 23H01182, and JST PRESTO Grant No.\ JPMJPR23F7, Japan. T.S. was supported by the JSPS fellowship Grant No.\ 23KJ0678.

\appendix

\section{Classical calculation of axion-induced electric field}
\label{appendix:classic}
\setcounter{equation}{0}

Consider the setup of qubits inside the conductor shield applied with static magnetic field $B_0$. Based on the Lagrangian given in Eq.\ \eqref{Lagrangian}, the equation of motion of the electromagnetic field inside the cavity reads
\begin{gather}
    \ddot{\vec{E}}-\nabla^2\vec{E}=-g_{a\gamma \gamma}m_a^2 a_0 \cos (m_a t-\alpha) \vec{B}_0,\\
    \vec{\nabla} \cdot \vec{E}=0 \label{eq:EOME}
\end{gather}
where we keep terms relevant for the case of $\vec{E}\sim O(g_{a\gamma \gamma})$. In addition, the electric field satisfies the following boundary condition
\begin{gather}
    \vec{E}|_\parallel=0,
\end{gather}
where $|_\parallel$ denotes the component of the field at the boundary of the shielding cavity and parallel to the wall of the shielding cavity.
We can obtain the solution by expanding the field $\vec{E}$ using cavity mode function $\vec{E}_m$:
\begin{equation}
    \vec{E}(\vec{x},t)= \sum_m c_m(t)\vec{E}_m(\vec{x})
\end{equation}
where mode functions satisfy Eqs.\ \eqref{Em1}, \eqref{Em2} and \eqref{Em3}, as well as the orthogonality condition given in Eq.\ \eqref{Em4}. 

The equation of motion of electric field then reads as
\begin{equation}
    \ddot{c}_m(t)-\omega_m^2 c_m=-g_{a\gamma \gamma} m_a^2 a_0 \cos (m_a t-\alpha)  \int d^3x \vec{E}_m \cdot \vec{B}_0,
\end{equation}
which give solution
\begin{equation}
    c_m(t)=g_{a\gamma \gamma } a_0 \cos (m_a t-\alpha) \frac{m_a^2}{m_a^2-\omega_m^2} \int d^3x \vec{E}_m \cdot \vec{B}_0.
\end{equation}
This gives the axion-induced electric field inside the cavity as
\begin{equation}
    \vec{E}(\vec{x},t)=g_{a\gamma \gamma } a_0 \cos (m_a t-\alpha) \sum_m \vec{E}_m (\vec{x}) 
    \frac{m_a^2}{m_a^2-\omega_m^2} \int d^3x' \vec{E}_m(\vec{x}') \cdot \vec{B}_0,
\end{equation}
which is consistent with the result given by Eq.~\eqref{eq:Esol}. For a similar derivation, see also Ref. \cite{Chaudhuri:2014dla}.

\section{Mode mixing in strong coupling regime} 
\label{appendix:strongcoupling}
\setcounter{equation}{0}

In this Appendix, we discuss the mixing between the cavity photon and qubit for the case that the qubit frequency is close to one of the cavity-mode frequencies in the strong coupling regime. In such a case, the mixing angle of those modes becomes sizable.

Considering the electromagnetic field inside the cavity, we have the Hamiltonian given by Eq.~\eqref{eq:H_Ea}.
On the other hand, based on Ref.\ \cite{Chen:2022quj} as well as Eq.~\eqref{eq:Eexpress}, effective Hamiltonian of the qubit coupled to the cavity photon is given by
\begin{align}
    H_{\rm q}=\omega_{\rm q} \ket{e} \bra{e}+\sum_m i\lambda_m (b_m - b^\dagger_m)(\ket{g}\bra{e}+\ket{e}\bra{g}),
    \label{eq:HqubitE}
\end{align}
where
\begin{align}
   \lambda_m \equiv \sqrt{\frac{\omega_{\rm q} C}{2}}  \left( \frac{\omega_m}{2} \right)^{1/2}( \hat{z}\cdot \vec{E}_m (\vec{x})) d.
\end{align}
The effective Hamiltonian describing the total system is given by the sum of ones given in Eqs.\ \eqref{eq:H_Ea} and \eqref{eq:HqubitE}. The second term in Eq.~\eqref{eq:HqubitE} represents the interaction between the electric field and the capacitor in the transmon assuming that the normal vector of the capacitor points in $z$ direction. With the typical transmon circuit parameter, $\lambda_m \sim O(1) \ \mathrm{MHz}$. 

Now, we consider the case that $\omega_{\rm q}$ is close to one of the cavity-mode frequencies (denoted as $\omega_c$, or $m=c$). Particularly, we consider the strong coupling regime, i.e., $\lambda_c \gg \tau^{-1}$ where $\tau$ is the coherence time of the system including cavity. Then, the qubit excitation is dominantly through the mixing with $c$-th caivty mode and hence we neglect cavity modes other than $m=c$. Applying the rotation wave approximation (with neglecting simultaneous annihilation or creation terms), we write Hamiltonian in the following form:
\begin{align}
    H=H_1+H_2,
\end{align}
where
\begin{align}
    H_1= &\,  \omega_c b_c^\dagger b_c +\omega_{\rm q}  \ket{e}\bra{e} +i\lambda_c (b_c\ket{e}\bra{g}-b_c^\dagger\ket{g}\bra{e}),\\
    H_2=&\, i g_c (b_c e^{i(m_at - \alpha)} - b_c^\dagger e^{-i(m_at - \alpha)}).
\end{align} 
The former is the Hamiltonian of the qubit and cavity photon system whereas the latter describes the interaction between the axion and cavity photons. 

\begin{figure}[t]
  \centering
\includegraphics[width=0.6 \linewidth]{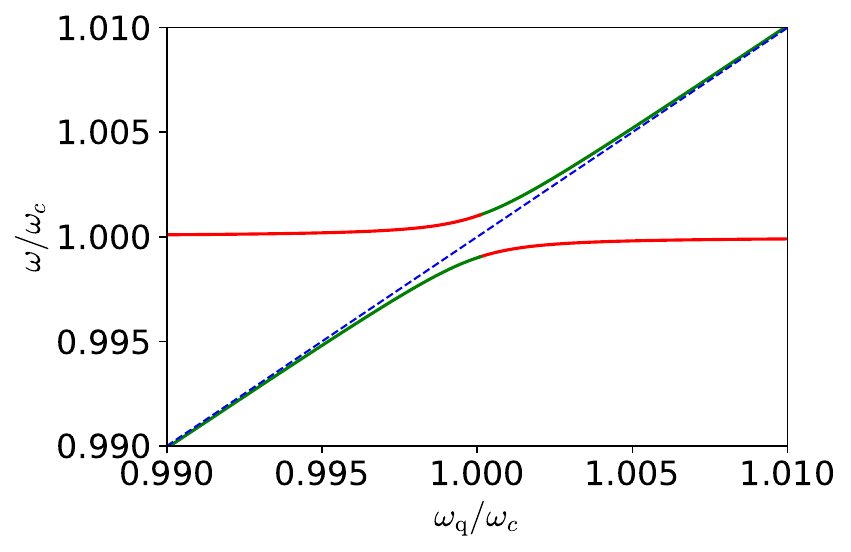}
  \caption{Spectrum structure when $\omega_{\rm q}$ near one of cavity mode with frequency $\omega_c$. 
  Spectrum $\omega_+$ and $\omega_-$ are contours above and below dashed blue line ($\omega=\omega_{\rm q}$), respectively. $\lambda_c/\omega_c=10^{-3}$ is assumed.
  The green contour is showing when state is dominated by qubit excitation covering $|m_a-\omega_c|>\lambda_c$, while the red contour is that dominated by cavity mode covering $|m_a-\omega_c|<\lambda_c$. }
  \label{fig:spectrum}
\end{figure}

In this case, we can work in the basis in which $H_1$ is diagonalized (see also Ref. \cite{walls2008quantum}). The Hamiltonian $H_1$ conserves the total number of excitation, defined as $\nu=\ket{e}\bra{e}+b_c^\dagger b_c$; for simplicity, we focus on the states with the excitation number being less than or equal to $1$ because we are interested in the case that the excitation probability is much smaller than $1$. Then we consider only $\ket{g}\otimes\ket{0}$, $\ket{g}\otimes b_c^\dagger\ket{0}$, and $\ket{e}\otimes\ket{0}$, where $\ket{0}$ is the cavity state satisfying $b_m\ket{0}=0$. The state $\ket{g}\otimes\ket{0}$ is the eigenstate of $H_1$, while the eigenstates with $\nu=1$ are given by
\begin{align}
  \begin{pmatrix}
    \ket{1}_+ \\ \ket{1}_-
  \end{pmatrix}
  \equiv P
  \begin{pmatrix}
    -i \ket{g}\otimes b_c^\dagger\ket{0} \\ \ket{e}\otimes \ket{0}
  \end{pmatrix},
\end{align} 
where 
\begin{align}
  P \equiv
  \begin{pmatrix}
    P_{+,1} & P_{+,2}\\
    P_{-,1} & P_{-,2}
  \end{pmatrix}=
  \left(
  \begin{array}{cc}
    \frac{r+\sqrt{r^2+4}}{\sqrt{\left(r+\sqrt{r^2+4}\right)^2+4}} & 
    \frac{2}{\sqrt{\left(r+\sqrt{r^2+4}\right)^2+4}}
    \\[2mm]
    \frac{r-\sqrt{r^2+4}}{\sqrt{\left(r-\sqrt{r^2+4}\right)^2+4}} &
    \frac{2}{\sqrt{\left(r-\sqrt{r^2+4}\right)^2+4}} \\
  \end{array}
  \right),
\end{align}
with $r\equiv(\omega_c-\omega_{\rm q})/\lambda_c$.  The eigenfrequencies in association with $\ket{1}_{\pm}$ are given by
\begin{equation}
    \omega_{\pm}=\frac{1}{2}\left( \omega_c+\omega_{\rm q}  \pm  \sqrt{(\omega_{\rm q}-\omega_c)^2+4  \lambda_c^2}\right).
\end{equation}
Note that qubit excitation state $\ket{e}\otimes\ket{0}$ becomes the major component of $\ket{1}_-$ and $\ket{1}_+$ for the case of $\omega_{\rm q} < \omega_c$ and $\omega_{\rm q} > \omega_c$, respectively. We also show the band structure in Fig.~\ref{fig:spectrum}, taking $\lambda_c/\omega_c=10^{-3}$.

\begin{figure}[t]
  \centering
\includegraphics[width=0.6 \linewidth]{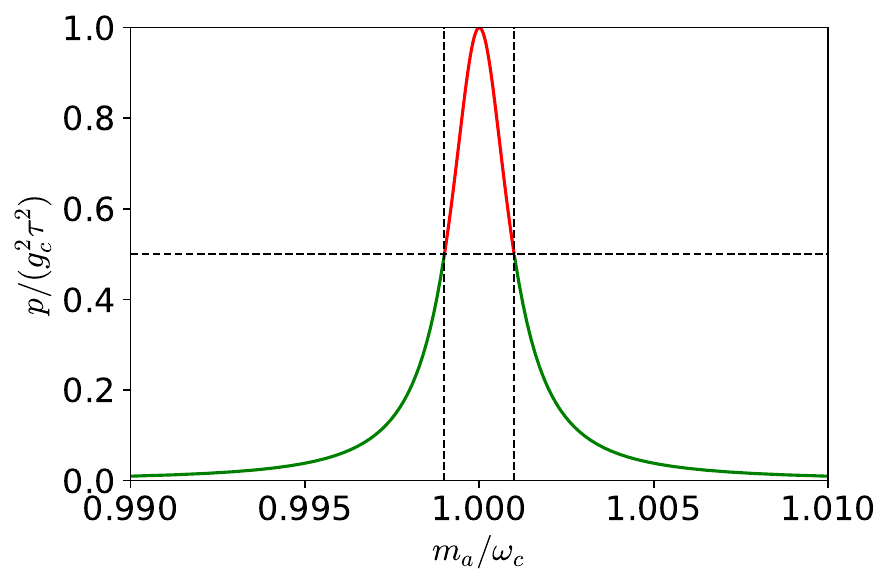}
  \caption{Excitation rate $p_{\pm}$ assuming $m_a=\omega_{\pm}$. The parameter relation $\lambda_c/\omega_c=10^{-3}$ is assumed. The green contour is the excitation of a state dominated by qubit excitation covering $|m_a-\omega_c|>\lambda_c$, while the red contour is that dominated by cavity mode covering $|m_a-\omega_c|<\lambda_c$.}
  \label{fig:testsignal}
\end{figure}

Now we consider the excitation to $\ket{1}_+$ or $\ket{1}_-$ from the ground state $\ket{g}\otimes\ket{0}$ in the limit of $m_a=\omega_{\pm}$; in such a resonance limit, the excitation probabilities to $\ket{1}_+$ and $\ket{1}_-$ are 
\begin{gather}
  p_{+}(\tau)=g_c^2 P_{+,1}^2 \tau^2,  \quad \mbox{and} \quad
  p_{-}(\tau)=g_c^2 P_{-,1}^2 \tau^2,
  \label{eq:pgemix}
\end{gather}
respectively. In Fig.\ \ref{fig:testsignal}, we show $p_{\pm}$ as a function of $m_a$, assuming $m_a=\omega_{\pm}$. It includes both the excitation of mode $\ket{1}_{-}$ and $\ket{1}_+$ following band structure given in Fig. \ref{fig:spectrum}. In Fig.~\ref{fig:testsignal}, those dominated by qubit state are plotted in green while those dominated by the cavity mode are plotted in red.

In principle, excitations to both $\ket{1}_+$ or $\ket{1}_-$ states can be the signal of axion DM. However, it is non-trivial to read out these states. Conventionally, the qubit states are read out by measuring the state-dependent dispersive shift of the frequency of the readout resonator \cite{Blais_2004}. In the case of the strong mode mixing, however, the dispersive shift may be suppressed because the $\omega_+$- and $\omega_-$-modes are given by the mixed state of $\ket{g}\otimes b_c^\dagger\ket{0}$ and $\ket{e}\otimes \ket{0}$. We expect that the dispersive shift is suppressed by the $P_{\pm,2}^2$ (which is the square of the mixing angle of the qubit to the hybridized state). We expect that the readout is possible if $P_{\pm,2}^2$ is sizable. In our estimate of the sensitivity, we assume that the readout is possible if $P_{\pm,2}^2\geq\frac{1}{2}$; thus, we may access the axion mass satisfying $|m_a-\omega_c|>\lambda_c$. In Fig.\ \ref{fig:k1cavityreac}, the sensitivity for the region $|m_a-\omega_c|<\lambda_c$ is not plotted; however, because $\lambda_c$ is so small that our treatment of the region of $|m_a-\omega_c|<\lambda_c$ does not (almost) affect the figure.

Notice that the argument in this Appendix gives the relevant procedure to estimate the transition rate when the qubit frequency becomes close to one of the cavity frequencies. If we follow the argument in Section \ref{sec:transmonex}, the transition probability would be arbitrarily large as $\omega_{\rm q}\simeq m_a\rightarrow\omega_c$ (see Eq.\ \eqref{eq:kappa}).  In such a limit, however, the argument of in Section \ref{sec:transmonex} is not applicable (even though it holds when $\omega_{\rm q}$ is away from the qubit frequencies) but we should use Eq.\ \eqref{eq:pgemix} to estimate the transition probability. Particularly, when $\omega_{\rm q}\rightarrow\omega_c$, $p_{\pm}(\tau)\rightarrow \frac{1}{2}g_c^2 \tau^2$. We also note that, for a range of $\omega_{\rm q}$, both Eq.\ \eqref{eq:pge} (with Eq.\ \eqref{eq:kappa}) and Eq.\ \eqref{eq:pgemix} are reliable; one can easily see that this is the case when 
\begin{equation}
\lambda_c \ll |\omega_{\rm q}-\omega_c| \ll \omega_{\rm c}.
\label{omegaq-omegac}
\end{equation}
In such a region, transition probability based on two estimations well agree.  In our numerical calculation, we use the formulas given in Eqs.~\eqref{eq:pge} and \eqref{eq:pgemix} to estimate the excitation rate of transmon to cover the entire frequency region of our interest, converting the formulas at a frequency satisfying the condition \eqref{omegaq-omegac}.

\bibliographystyle{jhep}
\bibliography{ref}

\providecommand{\href}[2]{#2}\begingroup\raggedright\begin{thebibliography}{10}

\bibitem{Chou:2023hcc}
A.~Chou et~al., \emph{{Quantum Sensors for High Energy Physics}},  11, 2023 [\href{https://arxiv.org/abs/2311.01930}{{\ttfamily 2311.01930}}].

\bibitem{Kennedy:2020bac}
C.J.~Kennedy, E.~Oelker, J.M.~Robinson, T.~Bothwell, D.~Kedar, W.R.~Milner et~al., \emph{{Precision Metrology Meets Cosmology: Improved Constraints on Ultralight Dark Matter from Atom-Cavity Frequency Comparisons}}, \href{https://doi.org/10.1103/PhysRevLett.125.201302}{\emph{Phys. Rev. Lett.} {\bfseries 125} (2020) 201302} [\href{https://arxiv.org/abs/2008.08773}{{\ttfamily 2008.08773}}].

\bibitem{Dixit:2020ymh}
A.V.~Dixit, S.~Chakram, K.~He, A.~Agrawal, R.K.~Naik, D.I.~Schuster et~al., \emph{{Searching for Dark Matter with a Superconducting Qubit}}, \href{https://doi.org/10.1103/PhysRevLett.126.141302}{\emph{Phys. Rev. Lett.} {\bfseries 126} (2021) 141302} [\href{https://arxiv.org/abs/2008.12231}{{\ttfamily 2008.12231}}].

\bibitem{Chen:2022quj}
S.~Chen, H.~Fukuda, T.~Inada, T.~Moroi, T.~Nitta and T.~Sichanugrist, \emph{{Detecting Hidden Photon Dark Matter Using the Direct Excitation of Transmon Qubits}}, \href{https://doi.org/10.1103/PhysRevLett.131.211001}{\emph{Phys. Rev. Lett.} {\bfseries 131} (2023) 211001} [\href{https://arxiv.org/abs/2212.03884}{{\ttfamily 2212.03884}}].

\bibitem{Fan:2022uwu}
X.~Fan, G.~Gabrielse, P.W.~Graham, R.~Harnik, T.G.~Myers, H.~Ramani et~al., \emph{{One-Electron Quantum Cyclotron as a Milli-eV Dark-Photon Detector}}, \href{https://doi.org/10.1103/PhysRevLett.129.261801}{\emph{Phys. Rev. Lett.} {\bfseries 129} (2022) 261801} [\href{https://arxiv.org/abs/2208.06519}{{\ttfamily 2208.06519}}].

\bibitem{Chigusa:2023hms}
S.~Chigusa, M.~Hazumi, E.D.~Herbschleb, N.~Mizuochi and K.~Nakayama, \emph{{Light Dark Matter Search with Nitrogen-Vacancy Centers in Diamonds}},  \href{https://arxiv.org/abs/2302.12756}{{\ttfamily 2302.12756}}.

\bibitem{Engelhardt:2023qjf}
G.~Engelhardt, A.~Bhoonah and W.V.~Liu, \emph{{Detecting axion dark matter with Rydberg atoms via induced electric dipole transitions}}, \href{https://doi.org/10.1103/PhysRevResearch.6.023017}{\emph{Phys. Rev. Res.} {\bfseries 6} (2024) 023017} [\href{https://arxiv.org/abs/2304.05863}{{\ttfamily 2304.05863}}].

\bibitem{Chen:2023swh}
S.~Chen, H.~Fukuda, T.~Inada, T.~Moroi, T.~Nitta and T.~Sichanugrist, \emph{{Quantum Enhancement in Dark Matter Detection with Quantum Computation}},  \href{https://arxiv.org/abs/2311.10413}{{\ttfamily 2311.10413}}.

\bibitem{Agrawal:2023umy}
A.~Agrawal, A.V.~Dixit, T.~Roy, S.~Chakram, K.~He, R.K.~Naik et~al., \emph{{Stimulated Emission of Signal Photons from Dark Matter Waves}}, \href{https://doi.org/10.1103/PhysRevLett.132.140801}{\emph{Phys. Rev. Lett.} {\bfseries 132} (2024) 140801} [\href{https://arxiv.org/abs/2305.03700}{{\ttfamily 2305.03700}}].

\bibitem{Ito:2023zhp}
A.~Ito, R.~Kitano, W.~Nakano and R.~Takai, \emph{{Quantum entanglement of ions for light dark matter detection}}, \href{https://doi.org/10.1007/JHEP02(2024)124}{\emph{JHEP} {\bfseries 02} (2024) 124} [\href{https://arxiv.org/abs/2311.11632}{{\ttfamily 2311.11632}}].

\bibitem{Braggio:2024xed}
C.~Braggio et~al., \emph{{Quantum-enhanced sensing of axion dark matter with a transmon-based single microwave photon counter}},  \href{https://arxiv.org/abs/2403.02321}{{\ttfamily 2403.02321}}.

\bibitem{Das:2022srn}
A.~Das, N.~Kurinsky and R.K.~Leane, \emph{{Dark Matter Induced Power in Quantum Devices}}, \href{https://doi.org/10.1103/PhysRevLett.132.121801}{\emph{Phys. Rev. Lett.} {\bfseries 132} (2024) 121801} [\href{https://arxiv.org/abs/2210.09313}{{\ttfamily 2210.09313}}].

\bibitem{Das:2024jdz}
A.~Das, N.~Kurinsky and R.K.~Leane, \emph{{Transmon Qubit Constraints on Dark Matter-Nucleon Scattering}},  \href{https://arxiv.org/abs/2405.00112}{{\ttfamily 2405.00112}}.

\bibitem{Moretti2024}
R.~Moretti, H.A.~Corti, D.~Labranca, F.~Ahrens, G.~Avallone, D.~Babusci et~al., \emph{Design and simulation of a transmon qubit chip for axion detection}, \href{https://doi.org/10.1109/TASC.2024.3350582}{\emph{IEEE Transactions on Applied Superconductivity} {\bfseries 34} (2024) 1}.

\bibitem{Lamoreaux:2013koa}
S.K.~Lamoreaux, K.A.~van Bibber, K.W.~Lehnert and G.~Carosi, \emph{{Analysis of single-photon and linear amplifier detectors for microwave cavity dark matter axion searches}}, \href{https://doi.org/10.1103/PhysRevD.88.035020}{\emph{Phys. Rev. D} {\bfseries 88} (2013) 035020} [\href{https://arxiv.org/abs/1306.3591}{{\ttfamily 1306.3591}}].

\bibitem{doi:10.1126/science.1104149}
V.~Giovannetti, S.~Lloyd and L.~Maccone, \emph{Quantum-enhanced measurements: Beating the standard quantum limit}, \href{https://doi.org/10.1126/science.1104149}{\emph{Science} {\bfseries 306} (2004) 1330}.

\bibitem{Kjaergaard2020}
M.~Kjaergaard, M.E.~Schwartz, J.~Braumüller, P.~Krantz, J.I.-J.~Wang, S.~Gustavsson et~al., \emph{Superconducting qubits: Current state of play}, \href{https://doi.org/https://doi.org/10.1146/annurev-conmatphys-031119-050605}{\emph{Annual Review of Condensed Matter Physics} {\bfseries 11} (2020) 369}.

\bibitem{Koch:2007hay}
J.~Koch, T.M.~Yu, J.~Gambetta, A.A.~Houck, D.I.~Schuster, J.~Majer et~al., \emph{{Charge-insensitive qubit design derived from the Cooper pair box}}, \href{https://doi.org/10.1103/physreva.76.042319}{\emph{Phys. Rev. A} {\bfseries 76} (2007) 042319} [\href{https://arxiv.org/abs/cond-mat/0703002}{{\ttfamily cond-mat/0703002}}].

\bibitem{Arias:2012az}
P.~Arias, D.~Cadamuro, M.~Goodsell, J.~Jaeckel, J.~Redondo and A.~Ringwald, \emph{{WISPy Cold Dark Matter}}, \href{https://doi.org/10.1088/1475-7516/2012/06/013}{\emph{JCAP} {\bfseries 06} (2012) 013} [\href{https://arxiv.org/abs/1201.5902}{{\ttfamily 1201.5902}}].

\bibitem{Krause:2021llk}
J.~Krause, C.~Dickel, E.~Vaal, M.~Vielmetter, J.~Feng, R.~Bounds et~al., \emph{{Magnetic Field Resilience of Three-Dimensional Transmons with Thin-Film Al/AlOx/Al Josephson Junctions Approaching 1 T}}, \href{https://doi.org/10.1103/PhysRevApplied.17.034032}{\emph{Phys. Rev. Applied} {\bfseries 17} (2022) 034032} [\href{https://arxiv.org/abs/2111.01115}{{\ttfamily 2111.01115}}].

\bibitem{Rokhinson:2012dp}
L.P.~Rokhinson, X.~Liu and J.K.~Furdyna, \emph{{Observation of the fractional ac Josephson effect: the signature of Majorana particles}}, \href{https://doi.org/10.1038/nphys2429}{\emph{Nature Phys.} {\bfseries 8} (2012) 795} [\href{https://arxiv.org/abs/1204.4212}{{\ttfamily 1204.4212}}].

\bibitem{Bauriedl2022}
L.~Bauriedl, C.~B{\"a}uml, L.~Fuchs, C.~Baumgartner, N.~Paulik, J.M.~Bauer et~al., \emph{Supercurrent diode effect and magnetochiral anisotropy in few-layer nbse2}, \href{https://doi.org/10.1038/s41467-022-31954-5}{\emph{Nature Communications} {\bfseries 13} (2022) 4266}.

\bibitem{Dvir2024}
T.~Dvir, A.~Zalic, E.H.~Fyhn, M.~Amundsen, T.~Taniguchi, K.~Watanabe et~al., \emph{Planar graphene-${\mathrm{nbse}}_{2}$ josephson junctions in a parallel magnetic field}, \href{https://doi.org/10.1103/PhysRevB.103.115401}{\emph{Phys. Rev. B} {\bfseries 103} (2021) 115401}.

\bibitem{Banerjee2018}
A.~Banerjee, P.~Lampen-Kelley, J.~Knolle, C.~Balz, A.A.~Aczel, B.~Winn et~al., \emph{Excitations in the field-induced quantum spin liquid state of $\alpha$-rucl3}, \href{https://doi.org/10.1038/s41535-018-0079-2}{\emph{npj Quantum Materials} {\bfseries 3} (2018) 8}.

\bibitem{_vila_2020}
J.~Ávila, E.~Prada, P.~San-Jose and R.~Aguado, \emph{Majorana oscillations and parity crossings in semiconductor nanowire-based transmon qubits}, \href{https://doi.org/10.1103/physrevresearch.2.033493}{\emph{Physical Review Research} {\bfseries 2} (2020) }.

\bibitem{Borcsok2019}
B.~B{\"o}rcs{\"o}k, S.~Komori, A.I.~Buzdin and J.W.A.~Robinson, \emph{Fraunhofer patterns in magnetic josephson junctions with non-uniform magnetic susceptibility}, \href{https://doi.org/10.1038/s41598-019-41764-3}{\emph{Scientific Reports} {\bfseries 9} (2019) 5616}.

\bibitem{Graaf2020}
S.E.~de~Graaf, L.~Faoro, L.B.~Ioffe, S.~Mahashabde, J.J.~Burnett, T.~Lindström et~al., \emph{Two-level systems in superconducting quantum devices due to trapped quasiparticles}, \href{https://doi.org/10.1126/sciadv.abc5055}{\emph{Science Advances} {\bfseries 6} (2020) eabc5055} [\href{https://arxiv.org/abs/https://www.science.org/doi/pdf/10.1126/sciadv.abc5055}{{\ttfamily https://www.science.org/doi/pdf/10.1126/sciadv.abc5055}}].

\bibitem{Kim:1979if}
J.E.~Kim, \emph{{Weak Interaction Singlet and Strong CP Invariance}}, \href{https://doi.org/10.1103/PhysRevLett.43.103}{\emph{Phys. Rev. Lett.} {\bfseries 43} (1979) 103}.

\bibitem{Shifman:1979if}
M.A.~Shifman, A.I.~Vainshtein and V.I.~Zakharov, \emph{{Can Confinement Ensure Natural CP Invariance of Strong Interactions?}}, \href{https://doi.org/10.1016/0550-3213(80)90209-6}{\emph{Nucl. Phys. B} {\bfseries 166} (1980) 493}.

\bibitem{Dine:1981rt}
M.~Dine, W.~Fischler and M.~Srednicki, \emph{{A Simple Solution to the Strong CP Problem with a Harmless Axion}}, \href{https://doi.org/10.1016/0370-2693(81)90590-6}{\emph{Phys. Lett. B} {\bfseries 104} (1981) 199}.

\bibitem{Zhitnitsky:1980tq}
A.R.~Zhitnitsky, \emph{{On Possible Suppression of the Axion Hadron Interactions. (In Russian)}}, {\emph{Sov. J. Nucl. Phys.} {\bfseries 31} (1980) 260}.

\bibitem{Peccei:1977hh}
R.D.~Peccei and H.R.~Quinn, \emph{{CP Conservation in the Presence of Instantons}}, \href{https://doi.org/10.1103/PhysRevLett.38.1440}{\emph{Phys. Rev. Lett.} {\bfseries 38} (1977) 1440}.

\bibitem{Peccei:1977ur}
R.D.~Peccei and H.R.~Quinn, \emph{{Constraints Imposed by CP Conservation in the Presence of Instantons}}, \href{https://doi.org/10.1103/PhysRevD.16.1791}{\emph{Phys. Rev. D} {\bfseries 16} (1977) 1791}.

\bibitem{Weinberg:1977ma}
S.~Weinberg, \emph{{A New Light Boson?}}, \href{https://doi.org/10.1103/PhysRevLett.40.223}{\emph{Phys. Rev. Lett.} {\bfseries 40} (1978) 223}.

\bibitem{Wilczek:1977pj}
F.~Wilczek, \emph{{Problem of Strong $P$ and $T$ Invariance in the Presence of Instantons}}, \href{https://doi.org/10.1103/PhysRevLett.40.279}{\emph{Phys. Rev. Lett.} {\bfseries 40} (1978) 279}.

\bibitem{Gorghetto:2018ocs}
M.~Gorghetto and G.~Villadoro, \emph{{Topological Susceptibility and QCD Axion Mass: QED and NNLO corrections}}, \href{https://doi.org/10.1007/JHEP03(2019)033}{\emph{JHEP} {\bfseries 03} (2019) 033} [\href{https://arxiv.org/abs/1812.01008}{{\ttfamily 1812.01008}}].

\bibitem{GrillidiCortona:2015jxo}
G.~Grilli~di Cortona, E.~Hardy, J.~Pardo~Vega and G.~Villadoro, \emph{{The QCD axion, precisely}}, \href{https://doi.org/10.1007/JHEP01(2016)034}{\emph{JHEP} {\bfseries 01} (2016) 034} [\href{https://arxiv.org/abs/1511.02867}{{\ttfamily 1511.02867}}].

\bibitem{ADMX:2018gho}
{\scshape ADMX} collaboration, \emph{{A Search for Invisible Axion Dark Matter with the Axion Dark Matter Experiment}}, \href{https://doi.org/10.1103/PhysRevLett.120.151301}{\emph{Phys. Rev. Lett.} {\bfseries 120} (2018) 151301} [\href{https://arxiv.org/abs/1804.05750}{{\ttfamily 1804.05750}}].

\bibitem{CAPP:2020utb}
{\scshape CAPP} collaboration, \emph{{First Results from an Axion Haloscope at CAPP around 10.7 $\mu$eV}}, \href{https://doi.org/10.1103/PhysRevLett.126.191802}{\emph{Phys. Rev. Lett.} {\bfseries 126} (2021) 191802} [\href{https://arxiv.org/abs/2012.10764}{{\ttfamily 2012.10764}}].

\bibitem{HAYSTAC:2023cam}
{\scshape HAYSTAC} collaboration, \emph{{New results from HAYSTAC\textquoteright{}s phase II operation with a squeezed state receiver}}, \href{https://doi.org/10.1103/PhysRevD.107.072007}{\emph{Phys. Rev. D} {\bfseries 107} (2023) 072007} [\href{https://arxiv.org/abs/2301.09721}{{\ttfamily 2301.09721}}].

\bibitem{QUAX:2023gop}
{\scshape QUAX} collaboration, \emph{{Search for galactic axions with a traveling wave parametric amplifier}}, \href{https://doi.org/10.1103/PhysRevD.108.062005}{\emph{Phys. Rev. D} {\bfseries 108} (2023) 062005} [\href{https://arxiv.org/abs/2304.07505}{{\ttfamily 2304.07505}}].

\bibitem{S_miller2018}
S.~Miller, \emph{A tunable 20 GHz transmon qubit in a 3D cavity}, Semester Thesis (2018).

\bibitem{Chen:2022frn}
L.~Chen et~al., \emph{{Transmon qubit readout fidelity at the threshold for quantum error correction without a quantum-limited amplifier}}, \href{https://doi.org/10.1038/s41534-023-00709-5}{\emph{npj Quantum Inf.} {\bfseries 9} (2023) 37} [\href{https://arxiv.org/abs/2208.05879}{{\ttfamily 2208.05879}}].

\bibitem{Fowler:2012hwn}
A.G.~Fowler, M.~Mariantoni, J.M.~Martinis and A.N.~Cleland, \emph{{Surface codes: Towards practical large-scale quantum computation}}, \href{https://doi.org/10.1103/physreva.86.032324}{\emph{Phys. Rev. A} {\bfseries 86} (2012) 032324} [\href{https://arxiv.org/abs/1208.0928}{{\ttfamily 1208.0928}}].

\bibitem{Bravyi:2023qpn}
S.~Bravyi, A.W.~Cross, J.M.~Gambetta, D.~Maslov, P.~Rall and T.J.~Yoder, \emph{{High-threshold and low-overhead fault-tolerant quantum memory}}, \href{https://doi.org/10.1038/s41586-024-07107-7}{\emph{Nature} {\bfseries 627} (2024) 778} [\href{https://arxiv.org/abs/2308.07915}{{\ttfamily 2308.07915}}].

\bibitem{Aad2012}
G.~Aad, T.~Abajyan, B.~Abbott, J.~Abdallah, S.~{Abdel Khalek}, A.~Abdelalim et~al., \emph{Observation of a new particle in the search for the standard model higgs boson with the atlas detector at the lhc}, \href{https://doi.org/https://doi.org/10.1016/j.physletb.2012.08.020}{\emph{Physics Letters B} {\bfseries 716} (2012) 1}.

\bibitem{Dolan:2022kul}
M.J.~Dolan, F.J.~Hiskens and R.R.~Volkas, \emph{{Advancing globular cluster constraints on the axion-photon coupling}}, \href{https://doi.org/10.1088/1475-7516/2022/10/096}{\emph{JCAP} {\bfseries 10} (2022) 096} [\href{https://arxiv.org/abs/2207.03102}{{\ttfamily 2207.03102}}].

\bibitem{AxionLimits}
C.~O'Hare, ``cajohare/axionlimits: Axionlimits.'' \url{https://cajohare.github.io/AxionLimits/}, July, 2020.
\newblock 10.5281/zenodo.3932430.

\bibitem{Kjaergaard:2019lmy}
M.~Kjaergaard, M.E.~Schwartz, J.~Braum\"uller, P.~Krantz, J.I.-J.~Wang, S.~Gustavsson et~al., \emph{{Superconducting Qubits: Current State of Play}}, \href{https://doi.org/10.1146/annurev-conmatphys-031119-050605}{\emph{Physics} {\bfseries 11} (2020) 369} [\href{https://arxiv.org/abs/1905.13641}{{\ttfamily 1905.13641}}].

\bibitem{Place:2021elq}
A.P.M.~Place et~al., \emph{{New material platform for superconducting transmon qubits with coherence times exceeding 0.3 milliseconds}}, \href{https://doi.org/10.1038/s41467-021-22030-5}{\emph{Nature Commun.} {\bfseries 12} (2021) 1779}.

\bibitem{Wang_2022}
C.~Wang, X.~Li, H.~Xu, Z.~Li, J.~Wang, Z.~Yang et~al., \emph{Towards practical quantum computers: transmon qubit with a lifetime approaching 0.5 milliseconds}, \href{https://doi.org/10.1038/s41534-021-00510-2}{\emph{npj Quantum Information} {\bfseries 8} (2022) }.

\bibitem{10.1119/1.16243}
D.M.~Greenberger, M.A.~Horne, A.~Shimony and A.~Zeilinger, \emph{{Bell’s theorem without inequalities}}, \href{https://doi.org/10.1119/1.16243}{\emph{American Journal of Physics} {\bfseries 58} (1990) 1131}.

\bibitem{Li:2023cla}
Z.~Li et~al., \emph{{Error per single-qubit gate below 10$^{-4}$ in a superconducting qubit}}, \href{https://doi.org/10.1038/s41534-023-00781-x}{\emph{npj Quantum Inf.} {\bfseries 9} (2023) 111} [\href{https://arxiv.org/abs/2302.08690}{{\ttfamily 2302.08690}}].

\bibitem{Li:2024xbj}
R.~Li, K.~Kubo, Y.~Ho, Z.~Yan, Y.~Nakamura and H.~Goto, \emph{{Realization of High-Fidelity CZ Gate based on a Double-Transmon Coupler}},  \href{https://arxiv.org/abs/2402.18926}{{\ttfamily 2402.18926}}.

\bibitem{Kono:2018rhn}
S.~Kono, K.~Koshino, Y.~Tabuchi, A.~Noguchi and Y.~Nakamura, \emph{{Quantum non-demolition detection of an itinerant microwave photon}}, \href{https://doi.org/10.1038/s41567-018-0066-3}{\emph{Nature Phys.} {\bfseries 14} (2018) 546}.

\bibitem{IBM}
IBM Quantum: \url{https://quantum-computing.ibm.com}.

\bibitem{ionQ}
IonQ Trapped Ion Quantum Computing: \url{https://ionq.com}.

\bibitem{Arute:2019zxq}
F.~Arute et~al., \emph{{Quantum supremacy using a programmable superconducting processor}}, \href{https://doi.org/10.1038/s41586-019-1666-5}{\emph{Nature} {\bfseries 574} (2019) 505} [\href{https://arxiv.org/abs/1910.11333}{{\ttfamily 1910.11333}}].

\bibitem{Chaudhuri:2014dla}
S.~Chaudhuri, P.W.~Graham, K.~Irwin, J.~Mardon, S.~Rajendran and Y.~Zhao, \emph{{Radio for hidden-photon dark matter detection}}, \href{https://doi.org/10.1103/PhysRevD.92.075012}{\emph{Phys. Rev. D} {\bfseries 92} (2015) 075012} [\href{https://arxiv.org/abs/1411.7382}{{\ttfamily 1411.7382}}].

\bibitem{walls2008quantum}
D.~Walls and G.~Milburn, \emph{Quantum Optics}, Springer Berlin Heidelberg (2008).

\bibitem{Blais_2004}
A.~Blais, R.-S.~Huang, A.~Wallraff, S.M.~Girvin and R.J.~Schoelkopf, \emph{Cavity quantum electrodynamics for superconducting electrical circuits: An architecture for quantum computation}, \href{https://doi.org/10.1103/physreva.69.062320}{\emph{Physical Review A} {\bfseries 69} (2004) }.

\end{thebibliography}\endgroup

\end{document}